\documentclass[sigconf,authorversion,nonacm,screen]{acmart}
\AtBeginDocument{%
  \providecommand\BibTeX{{%
    \normalfont B\kern-0.5em{\scshape i\kern-0.25em b}\kern-0.8em\TeX}}}

\usepackage{soul}
\usepackage{svg}
\usepackage{CJKutf8}
\usepackage{booktabs}
\usepackage{multirow}
\usepackage{siunitx}
\usepackage{makecell}
\usepackage{balance}

\copyrightyear{2024}
\acmYear{2024}
\setcopyright{acmlicensed}\acmConference[CHI '24]{Proceedings of the CHI Conference on Human Factors in Computing Systems}{May 11--16, 2024}{Honolulu, HI, USA}
\acmBooktitle{Proceedings of the CHI Conference on Human Factors in Computing Systems (CHI '24), May 11--16, 2024, Honolulu, HI, USA}
\acmDOI{10.1145/3613904.3642303}
\acmISBN{979-8-4007-0330-0/24/05}

\begin{document}
\begin{CJK}{UTF8}{ipxm}

\title[Silver-Tongued and Sundry]{Silver-Tongued and Sundry:\\Exploring Intersectional Pronouns with ChatGPT}

\author{Takao Fujii}
\email{fujii.t.av@m.titech.ac.jp}
\orcid{0009-0004-5059-3323}
\affiliation{%
  \institution{Tokyo Institute of Technology}
  \city{Tokyo}
  \country{Japan}
}

\author{Katie Seaborn}
\email{seaborn.k.aa@m.titech.ac.jp}
\orcid{0000-0002-7812-9096}
\affiliation{%
  \institution{Tokyo Institute of Technology}
  \city{Tokyo}
  \country{Japan}
}

\author{Madeleine Steeds}
\orcid{0000-0003-3767-292X}
\email{madeleine.steeds@ucd.ie}
\affiliation{%
  \institution{University College Dublin}
  \city{Dublin}
  \country{Ireland}
}

\renewcommand{\shortauthors}{Fujii, Seaborn, and Steeds}

\begin{abstract}
  ChatGPT is a conversational agent built on a large language model. Trained on a significant portion of human output, ChatGPT can mimic people to a degree. As such, we need to consider what social identities ChatGPT simulates (or can be designed to simulate). In this study, we explored the case of identity simulation through Japanese first-person pronouns, which are tightly connected to social identities in intersectional ways, i.e., intersectional pronouns. We conducted a controlled online experiment where people from two regions in Japan (Kanto and Kinki) witnessed interactions with ChatGPT using ten sets of first-person pronouns. We discovered that pronouns alone can evoke perceptions of social identities in ChatGPT at the intersections of gender, age, region, and formality, with caveats. This work highlights the importance of pronoun use for social identity simulation, provides a language-based methodology for culturally-sensitive persona development, and advances the potential of intersectional identities in intelligent agents.
\end{abstract}





\maketitle

\makeatletter \gdef\@ACM@checkaffil{} \makeatother


\section{Introduction}





The ChatGPT chatbot has become ubiquitous around the world, made possible through large language models (LLMs), specifically generative pre-trained transformer (GPT) models, trained on a gigantic human-generated data set made up of content from across the Internet\footnote{\url{https://help.openai.com/en/articles/6783457-what-is-chatgpt}}. Chats, forums, and message boards are just a few of the data forms responsible for ChatGPT's ability to converse naturally in a variety of languages. ChatGPT is so ``fluent'' that we may think of it as humanlike, even if we know better~\cite{nass1997machines,marchandot2023chatgpt}. Part of this humanlikeness is its use of \emph{social identity} cues, such as referring to itself using ``I,'' ``me,'' and ``my'' in English, i.e., using \emph{first-person pronouns}~\cite{loos2023using, balmer2023sociological, casal2023can}. But this raises ethical issues when leveraging ChatGPT and similar LLMs in the design of intelligent interactive agents.
Data sets used to train LLMs and intelligent agents are subject to biases~\cite{buolamwini2018gender, bolukbasi2016man}, including in the presence and use of social identity markers and notably pronouns \cite{seaborn2023transc,seaborn2023imlost}. Developers unconsciously embed mental models of the human world, including stereotypes, in their designs~\cite{friedman1996bias,wessel2022gender,lee2016robot,bergen2016d}. Users perceive humanlike cues in these artifacts that often reflect oversimplified models about people~\cite{nag2020gender,tay2014stereotypes,perugia2023models,eyssel2012s}. This renders technology as a means of reinforcing human stereotypes.
Many have investigated how agents equipped with AI shape perceptions through voice~\cite{Seaborn2021Voice}, speech patterns~\cite{tay2014stereotypes}, names~\cite{nag2020gender}, and other features linked to identity expression~\cite{rato2021towards}. Notably, pronouns are easily manipulated in machine-generated text~\cite{SeabornFrank2022Pepper,sun2019mitigating,adam2021ai,seaborn2023transc,seaborn2023imlost,cho2019measuring} and may indicate biases in the data sets or training methods \cite{sun2019mitigating,seaborn2023imlost,cho2019measuring}. 
As such, while pronouns can express identities and diversity, they may also be (un)consciously embodied by machines in the service of prejudice.


Pronouns are often associated with \emph{gender} in many languages. In English, for example, first-person pronouns like ``I'' and second-person pronouns like ``you'' are not gendered, but third-person pronouns like ``she'' are, creating a power imbalance in gender representation and expression~\cite{cameron2003language, nakamura2007language}. However, the Japanese language presents a different case: first-person pronouns \emph{are} gendered, in principal. While the gender-neutral 私 (watashi-kanji) exists, its use can be perceived as evasive, depending on the social context. Indeed, societal and state power structures enforce certain pronouns for certain people.
In elementary school, the use of\phantom{ }ぼく (boku) for boys and\phantom{ }わたし (watashi) for girls is imposed by teachers, textbooks, and the media~\cite{hirano1994heterosexism, maree2007language, nakamura2022feminist}.
Using pronouns deemed ``incorrect'' for one's gender is considered a form of rule-breaking, to the extent that young feminist women, girls, and femmes began using\phantom{ }ぼく (boku) to protest legislated gender norms~\cite{nakamura2007language, miyazaki2016japanese}. This dates back to the Meiji era (1868-1912), when the government began regulating the use of language~\cite{nakamura2007language}. Gender and sexual minorities, including trans children, struggle with the enforcement of normative and binary first-person pronoun use instilled upon them from childhood onwards~\cite{hirano1994heterosexism, maree2007language}. All in all, Japanese first-person pronouns are powerful identity markers, socially sensitive, and implicated by individual and institutional power.

Japanese first-person pronouns are also \emph{intersectional}, marking gender but also region, age, class, and formality~\cite{nakamura2022feminist, nakamura2014gender}. In short, gender and sundry\footnote{We have taken a cue from the phrase ``all and sundry,'' meaning ``everyone.'' We use ``sundry'' to mean variety, diversity, and plurality, as well as ``everything else,'' i.e., recognition of intersectional factors beyond gender.}. We can thus refer to Japanese first-person pronouns as \textbf{intersectional pronouns} or \textbf{\underline{交差代名詞} (kousa-daimeishi)} in Japanese. Intersectionality theory~\cite{crenshaw2013mapping, collins2022black} explains and provides an analytical framework for understanding how power operates through this plurality of social identities. Saying ``I'' in Japanese means choosing from a range of pronouns that express gender \emph{as well as} one or more social identities to others. This is a double-edged sword. People can immediately recognize complex archetypes and stereotypes based solely on the range of social identity markers associated with certain pronouns~\cite{nakamura2007language, nakamura2022feminist}.
However, media reliance on pronoun stereotypes has led to discrimination and prejudice even while enabling recognizable ``personas'' that express legitimate identities~\cite{kobayashi2013language, kinsui2003japanese, nakamura2007language}.
Given this, Japanese first-person pronouns must be analyzed in intersectional ways and approached with sensitivity to social power and stereotypes~\cite{nakamura2022feminist, ide1997women}. The rapid uptake of LLM-powered intelligent agents that speak ``like people'' and use first-person pronouns, such as ChatGPT, necessitates that we uncover how pronouns are---or could be---employed and whether this leads to perceptions of agent social identity that reify---or disrupt---stereotypes. While this may be especially pertinent to the Japanese socio-linguistic context, it reflects long-standing calls in HCI to more deeply engage with social identities in an intersectional way \cite{schlesinger2017intersectional, vieweg2015between, kumar2019intersectional}, especially beyond gender.

To this end, we explored ten sets of Japanese first-person pronouns as social identity cues for ChatGPT. Using mixed methods, we analyzed perceptions of ChatGPT's gender, age, region, and formality across two distinct regions in Japan (Kanto and Kinki). Our overarching research question (RQ) was: \textit{\textbf{Can ChatGPT elicit perceptions of intersectional social identities through the use of Japanese first-person pronouns?}} We contribute the following:

\begin{itemize}
    \item Empirical evidence that first-person pronouns alone can elicit perceptions of simple (i.e., gendered) and complex (i.e., intersectional) social identities in ChatGPT and likely other LLM-powered intelligent agents
    \item Empirical evidence that regional variations may require social identity markers beyond first-person pronouns
    \item A simple and quick methodology for crafting gendered and intersectional personas using LLM-based intelligent agents like ChatGPT for Japanese user groups
\end{itemize}

We believe this work is the first to explore the use of first-person pronouns for eliciting perceptions of intersectional social identities in LLM-based intelligent agents like ChatGPT. We also offer a deeper approach that features regional comparisons between the west (Kinki) and east (Kanto) of Japan. We aim to spark work on Japanese-speaking intelligent agents in an ethical way that centres social identity, as well as shift HCI research away from the current WEIRD (Western, Educated, Industrial, Rich, Democratic) focus ~\cite{linxen2021weird} to other techno-social contexts implicated by LLM-based technologies.

\section{Background and Theory}

\subsection{Intersectionality and Identity in HCI}

The notion of identity is not new to the field of HCI. Personas~\cite{marsden2019personas}, participant demographics~\cite{schlesinger2017intersectional}, and, more recently, positionality~\cite{rode2011theoretical, alvarado2021decolonial} feature in work stretching back to our field's humble beginnings, notably prompted by feminist HCI scholarship~\cite{rode2011theoretical, bardzell2010feministhci}. Nevertheless, the nature of identity as \emph{intersectional} and how to acknowledge this element in design and research practice---and perhaps during meta-research and reflexivity---remains less explored~\cite{schlesinger2017intersectional, marsden2019personas, alvarado2021decolonial}. In their survey of the field, \citet{schlesinger2017intersectional} found that most research focused on one aspect of identity in isolation, obfuscating the intersectional nature of identities, with a bias towards gender and class. \citet{marsden2019personas} considered the case of personas, recognizing a practical trade-off when it comes to matters of identity: a desire for a simple tool at the cost of oversimplifying the true complexity of social identities for a given persona. \citet{kumar2019intersectional}, in motivating the area of intersectional computing, raised the issue of a narrow focus on how intersectionality has been constructed within HCI discourse so far: gender, race, and class. They pointed out that other features of identity, perhaps those more subtle or marginalized---including elements related to language---still need to be brought into the conversation. More subtle yet, the discourse on intersectionality has been embedded within a Western frame, perhaps owing to its roots in the American context~\cite{crenshaw2013mapping, collins2022black}, as well as the WEIRD backdrop of most participant research~\cite{henrich2010weirdest}, including at CHI~\cite{linxen2021weird}. Here, we take steps towards broadening the purview of intersectional identity work in HCI through the novel frame of Japanese intersectional identities as reified in first-person pronoun use by a sophisticated intelligent agent, i.e., ChatGPT.

\subsection{Intersectionality and Identity in AI and Intelligent Agents}

Identity is 
core to the ``human'' side of the HCI equation. But technology can also be perceived as having 
identities or characteristics related to human identities~\cite{nass1997machines}, even when artificial intelligence is not involved~\cite{steeds2021device}. Thus, it is not surprising that intelligent agents can be interpreted as social entities~\cite{purington2017alexa, nass1997machines} and ascribed human identities such as gender~\cite{martin2023hey, purington2017alexa}. This can have practical effects. Anthropomorphization, for example, may improve user satisfaction with technology~\cite{purington2017alexa}, impact likability~\cite{ernst2020impact} and influence user behaviour, such as by increasing purchase intent~\cite{martin2023hey}.

Still, most research has not been intersectional, notably focusing on gender in isolation over a plurality of intersecting identity factors~\cite{ciston2019intersectional,Jarrell2021usinginter,marti2023speculating}, at least when it comes to how intelligent agents may embody such identities. 
Moreover, most of the work has focused on the system side rather than on the agent side. For example, in their landmark paper, \citet{buolamwini2018gender} considered the cross-section of gender and race in AI-powered automated facial recognition systems, finding that the classifier for darker-skinned women-presenting subjects performed the worst. In another example, \citet{karizat2021algorithmic} explored AI-powered recommender algorithms, finding biases related to the system promoting or suppressing content based on the inferred intersectional identities of users.
Work on intersectionality and identity in intelligent agents is scant. In one exception, \citet{Jarrell2021usinginter} explored whether non-playable characters representing combinations of races and genders influenced, even unconsciously, racial and gender biases in an unrelated follow-up task. Like \citet{buolamwini2018gender}, they found that job applications from mock applicants representing Black feminine identities experienced the greatest levels of discrimination, although this was reduced if participants had previously played the game with a character embodying the same characteristics. A final note is that these examples focused on the \emph{visual} medium over other modalities of identity, such as language and pronouns. Still, they attest to the power of multiple interwoven social identities embodied in intelligent systems and agents.

When it comes to user perception of intersectional identities in intelligent agents, the \emph{act of attribution} is key. Intelligent agents such as ChatGPT do not \emph{have} identities the way people do~\cite{tay2014stereotypes,perugia2023models,edwards2019evaluations}. Instead, people \emph{perceive} identity markers \emph{elicited by} simulated identity performances---in our case, anchored to first-person pronoun use. Recognizing this, we will use such phrasing as participant ``gendering,'' ChatGPT ``genderedness,'' and ``perceived [gender].''

\subsection{Identity and Japanese First-person Pronouns}

First-person pronouns in Japanese express simple, i.e., gender alone, and complex, i.e., multiple and often intersecting, social identities~\cite{nakamura2022feminist, nakamura2007language, nakamura2014gender, maree2007language, kinsui2003japanese, yee2021japanese}. Gender may be salient, even while pronouns are marked by other identities. As Nakamura explains for the Japanese context, ``gender is considered to constitute only one facet of one's identity, along with other facets including class, age, region, ethnicity, sexuality, and social role''~\cite[p.~278]{nakamura2022feminist}.
We could not find any work that has explicitly considered Japanese pronoun use by intelligent agents. Still, pronouns are used to express social identities, especially gender, in anime characters~\cite{hiramoto2013hey} and translated dialogue~\cite{nakamura2020formation}. Whether this works and to what extent for \emph{interactive and imperfect intelligent agents} remains unknown. ChatGPT is an ideal case because of the strength of its underlying language models and perceived fluency (cf. \citet{marchandot2023chatgpt}). This leads us to ask \textbf{\textit{RQ1: Do people's perceptions of ChatGPT's gender change based on first-person pronoun use?}} We can make predictions based on the identities linked to each pronoun, explained next.

The most common first-person pronoun is\phantom{ }私 (watashi-kanji), which is used by people of all genders in public situations and may be considered the semantic prime~\cite{yee2021japanese}. It is also written and pronounced as\phantom{ }わたし (watashi) and\phantom{ }あたし (atashi).
Women use these alternatives in private, with\phantom{ }あたし (atashi) being more feminine~\cite{nakamura2022feminist, miyazaki2016japanese, mcgloin1990aspects}. Men use\phantom{ }ぼく (boku) and\phantom{ }おれ (ore), with\phantom{ }おれ (ore) considered more masculine~\cite{miyazaki2016japanese, mcgloin1990aspects}. Given the default setting, we first examined whether ChatGPT using\phantom{ }私 (watashi-kanji) was perceived as gender-neutral (H1.1). Then we investigated whether ChatGPT using\phantom{ }わたし (watashi) and\phantom{ }あたし (atashi) were perceived as feminine (H1.2, H1.3) and\phantom{ }ぼく (boku) and\phantom{ }おれ (ore) as masculine (H1.4, H1.5). We also compared the most feminine and masculine pronouns,\phantom{ }あたし (atashi) and\phantom{ }おれ (ore), to their less gendered counterparts,\phantom{ }わたし (watashi) and\phantom{ }ぼく (boku), to confirm the degree of gendering (H1.6, H1.7).

\begin{itemize}
    \item H1.1: ChatGPT using\phantom{ }私 (watashi-kanji) will be perceived as gender-neutral.
    \item H1.2: ChatGPT using\phantom{ }わたし (watashi) will be perceived as feminine.
    \item H1.3: ChatGPT using\phantom{ }あたし (atashi) will be perceived as feminine.
    \item H1.4: ChatGPT using\phantom{ }おれ  (ore) will be perceived as masculine.
    \item H1.5: ChatGPT using\phantom{ }ぼく (boku) will be perceived as masculine.
    \item H1.6: ChatGPT using\phantom{ }あたし (atashi) will be perceived as more feminine than\phantom{ }わたし (watashi).
    \item H1.7: ChatGPT using\phantom{ }おれ (ore) will be perceived as more masculine than\phantom{ }ぼく (boku).
\end{itemize}

Many Japanese first-person pronouns express not only gender but other social identities, by degrees of formality and with regional variations~\cite{ide1997women,kumadaki2006language,nakamura2007language, Ogino2007jibun}. This leads us to ask: \textbf{\textit{RQ2: Do people's perceptions of ChatGPT's identity change based on intersectional first-person pronoun use, especially gender and sundry: age, region, and formality (National Sundry)?}} In formal situations, other pronunciations of the kanji\phantom{ }私 are used, such as\phantom{ }わたくし (watakushi) and\phantom{ }あたくし (atakushi)~\cite{mcgloin1990aspects, kumadaki2006language}.\phantom{ }あたくし (atakushi) is considered feminine~\cite{nakamura2007language}.\phantom{ }じぶん (jibun) is a masculine pronoun derived from military culture~\cite{kinuhata2007guntai}, but is sometimes used by young men in formal situations~\cite{Ogino2007jibun}. These formal pronouns are considered more urban, adult, and high status. Dialect is also crucial. In rural areas, \phantom{ }うち (uchi) is used by women, but recently urban female students who deviate from the femininity of\phantom{ }わたし (watashi) and\phantom{ }あたし (atashi) also use it~\cite{nakamura2007language, miyazaki2016japanese, Ogino2007jibun}. Also,\phantom{ }わし (washi) invokes the image of an older adult man throughout Japan, but in rural areas it has long used by both men and women~\cite{nakamura2007language, kumagai2010dialect}. Thus, first-person pronouns in Japan are truly \textit{sundry}, with each expressing complex and contextual cross-sectional identities.

\begin{itemize}
    \item H2.1: ChatGPT using\phantom{ }わたくし (watakushi) will be perceived as gender-neutral, not young, urban, and formal. 
    \item H2.2: ChatGPT using\phantom{ }あたくし (atakushi) will be perceived as feminine, not young,urban, and formal.
    \item H2.3: ChatGPT using\phantom{ }うち (uchi) will be perceived as feminine, ageless, regionless, and casual.
    \item H2.4: ChatGPT using\phantom{ }じぶん (jibun) will be perceived as masculine, ageless, urban, and formal.
    \item H2.5: ChatGPT using\phantom{ }わし (washi) will be perceived as masculine, older, rural, and casual.
\end{itemize}

In urban areas, people tend not to use dialects and are susceptible to standardization through media and policies more so than those in rural areas~\cite{nakamura2022feminist, nakamura2007language, kinsui2003japanese, kumagai2010dialect, kobayashi2013language}. As such, urban dwellers may hold stronger gender, age, and/or regional stereotypes about pronouns.
So, we also asked \textbf{\textit{RQ3: Do people's perceptions of ChatGPT's gender change based on regional differences in first-person pronoun use (Local Sundry)?}} In this study, we focused on two major regions in Japan: Kanto---an urban area---and Kinki---a rural area. In urban areas (Kanto),\phantom{ }うち (uchi) may be perceived as less feminine than\phantom{ }わたし(watashi) and\phantom{ }あたし (atashi) due to girls not commonly using\phantom{ }うち (uchi) alongside standardization (H3.1). In rural areas (Kinki),\phantom{ }わし (washi) is perceived as less masculine than\phantom{ }ぼく (boku) and\phantom{ }おれ (ore), due to older women using\phantom{ }わし (washi) alongside the low influence of standardization (H3.2).

\begin{itemize}
    \item H3.1: ChatGPT using\phantom{ }わし (washi) will be perceived as less masculine than\phantom{ }ぼく (boku) and\phantom{ }おれ (ore) in Kinki, which is rural, compared to Kanto, which is urban.
    \item H3.2: ChatGPT using\phantom{ }うち (uchi) will be perceived as less feminine than\phantom{ }わたし (watashi) and\phantom{ }あたし (atashi) in Kanto, which is urban, compared to Kinki, which is rural.
\end{itemize}

\subsection{Theoretical Model: Mapping First-Person Pronouns to Factors of Social Identity, Intersectionally}

\begin{figure*}[htbp]
  \centering
  \includegraphics[width=.75\textwidth]{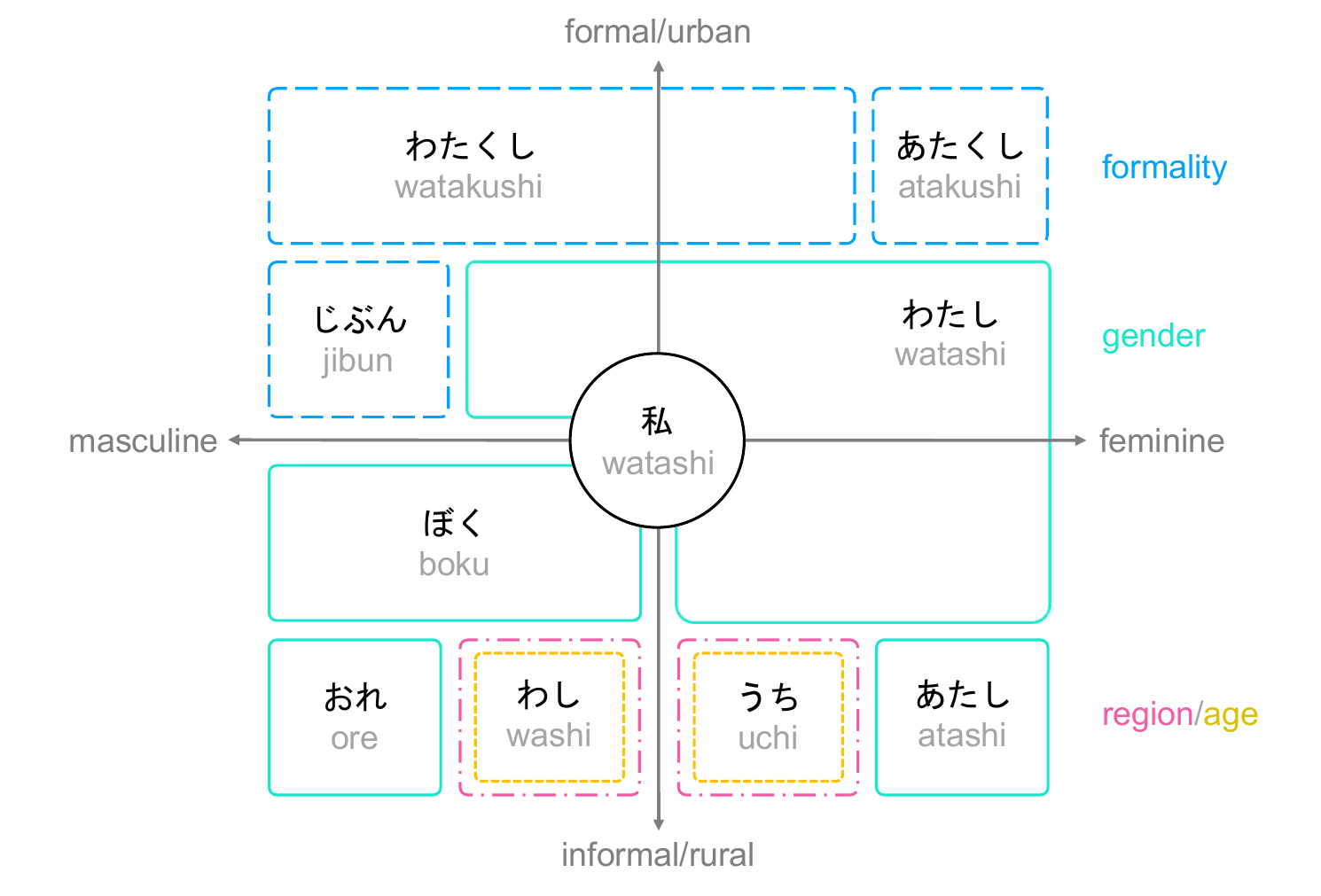}
  \caption{Pronoun map with the ten types of Japanese first-person pronouns intersectionally linked to social identities in Japan. Blue dashed lines: Formality. Green solid lines: Gender. Pink dashed lines with a dot: Region. Gold dotted lines: Age.}
  \Description{Pronoun map with two axes: formality/region/age and gender, with 私 (watashi-kanji) as the neutral centre.}
  \label{fig:pronounmap}
\end{figure*}

Our theoretical model of how Japanese first-person pronouns map onto social identities in Japan is presented in Figure~\ref{fig:pronounmap}. We can place\phantom{ }私 (watashi-kanji) as the neutral centre or prime. One axis represents gender, from masculine to feminine, which implicates all pronouns. The other axis represents the intersection of formality and region, with more formal and urban pronouns at one end, and more informal and rural pronouns at the other. Notably, age also implicates the informal/rural pronouns, with\phantom{ }わし (washi) used by older rural men in informal settings and\phantom{ }うち (uchi) used by young rural women in informal settings. 
Next, we describe how we explored these predictions experimentally with ChatGPT.

\section{Methods}
We ran an online experiment with one within factor (ten sets of Japanese pronouns used by ChatGPT) and one between factor (participant region: Kanto or Kinki). Our protocol was registered before data collection on August 11\textsuperscript{th}, 2023\footnote{\url{https://osf.io/qbhsj}; an earlier draft was accidentally registered on June 23\textsuperscript{rd}, 2023}. This research was approved by the Tokyo Institute of Technology ethics board (\#2023081).

\subsection{Participants}
201 valid responses (Kanto n=99, Kinki n=102; n=9 empty removed; refer to Table~\ref{tab:demographics}) were gathered via Yahoo! Crowdsourcing between August 11--12\textsuperscript{th}, 2023 from 210 people who were paid 227 yen for $\sim$15 mins. Yahoo! Crowdsourcing is a Japanese online recruitment platform that guarantees unique respondents with certain demographics, i.e., gender and region, through account verification, all while maintaining anonymity\footnote{\url{https://crowdsourcing.yahoo.co.jp/} (Japanese)}. Exact statistics are elusive, but Yahoo! services are among the most accessed in Japan; for instance, 2018 statistics show that it was the most popular provider, with 67.43 million active users\footnote{\url{https://www.statista.com/statistics/1071192/japan-most-popular-websites-online-services-by-monthly-active-users/}}, suggesting a wide reach. Yahoo! Crowdsourcing is used by researchers in Japan (e.g., \cite{to2016crowd,seaborn_can_2023}) and may be considered equivalent to online crowdsourcing platforms like Amazon Mechanical Turk and Prolific, i.e., workers sign up for a task and receive payment on completion. Unlike these services, Yahoo! Crowdsourcing requires a verified ID, such as a driver's license or the Japanese My Number identity card. Still, as far as we know, there has been no audit or quality assessment for research purposes. As such, we implemented an attention check (refer to~\ref{section:procedure}).

\begin{table*}
  \caption{Participant demographics. Empty cells equate to zero. Note: Multiple options could be selected for gender. 
  }
  \label{tab:demographics}
  \begin{tabular}{lllll}
    \toprule
    Demographic &Option &Kanto &Kinki &Total\\
    \midrule
    Gender &Man &43 &36 &79 \\
     &Woman &55 &66 &121 \\
     &Non-binary/X-gender & & & \\
     &Transgender & & & \\
     &Another gender or N/A &1 & &1 \\
     \midrule
    Age &18-24 &4 & &4\\
     &25-34 &10 &12 &22\\
     &35-44 &25 &25 &50\\
     &45-54 &39 &34 &73\\
     &55-64 &17 &19 &36\\
     &65-74 &2 &8 &10\\
     &75+ &2 &3 &5\\
     \midrule
    Education &Less than high school &3 2& &5\\
     &Technical high school or junior college &16 &23 &39\\
     &HS or equivalent &28 &26 &54\\
     &Bachelors &43 &39 &82\\
     &Graduate degree &6 &8 &14\\
     &Other or N/A &3 &4 &7\\
     \midrule
    Chat AI use &Daily &1 & &1\\
     &Multiple times a week &13 &5 &18\\
     &Once a week &9 &12 &21\\
     &Monthly &13 &14 &27\\
     &Never &63 &69 &132\\
     &Used to but stopped & & &\\
    \bottomrule
  \end{tabular}
\end{table*}

\label{section:procedure}
\subsection{Procedure}
Participants were informed about the study on the recruiting platform (Yahoo! Crowdsourcing). They were asked to choose one of three links to each scenario based on their date of birth (1\textsuperscript{st}-31\textsuperscript{st}) to avoid novelty and order effects~\cite{schuman_questions_1996}. First, they were asked to provide consent. The next page was an attention check, where they were asked to watch a video and input the number generated in a short conversation with ChatGPT that did not feature any first-person pronouns. Next, they were presented with ten pages for each of the ten pronoun types in random order. On each page, they were directed to watch a brief video\phantom{ }(こちらの動画を視聴してください) depicting a user interaction with ChatGPT using the first-person pronouns. They were informed there was no audio\phantom{ }(音声はありません) and that they could watch the video as many times as they liked\phantom{ }(何度でも自由に視聴し直してもらって構いません). They were then asked to provide their impressions of "the AI" through rating scale and open-ended items. At the end of the study, participants provided their demographics. They then received a code for payment on Yahoo! Crowdsourcing.

\subsection{Materials}
\begin{figure}[!tbh]
\centering
\begin{minipage}[b]{0.53\textwidth}
    \includegraphics[width=\textwidth]{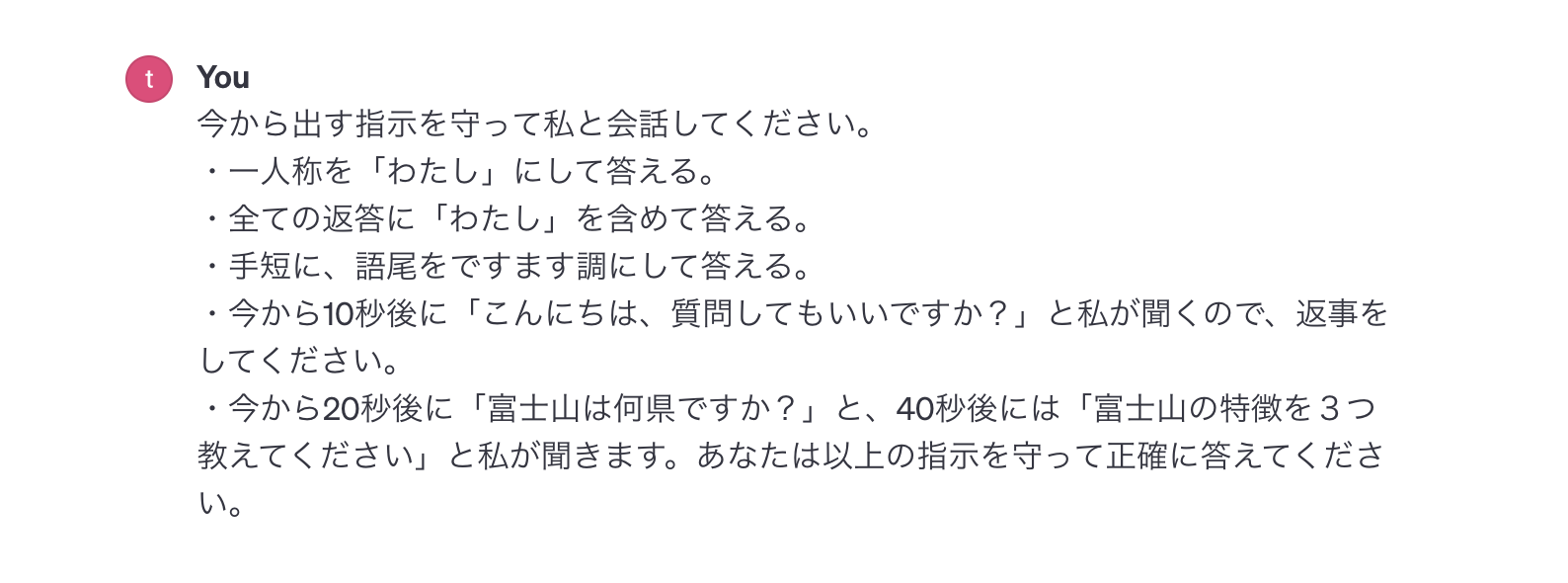}
    \caption{Example with all five ChatGPT prompts.}
    \label{fig:prompt}
    \Description{Example of the ChatGPT prompts used to set the type of first-person pronouns, use these pronouns, answer with neutral forms of verbs, trigger to start of the conversation, and set the topic of the conversation.}
\end{minipage}
\hfill
\begin{minipage}[b]{0.45\textwidth}
    \includegraphics[width=\textwidth]{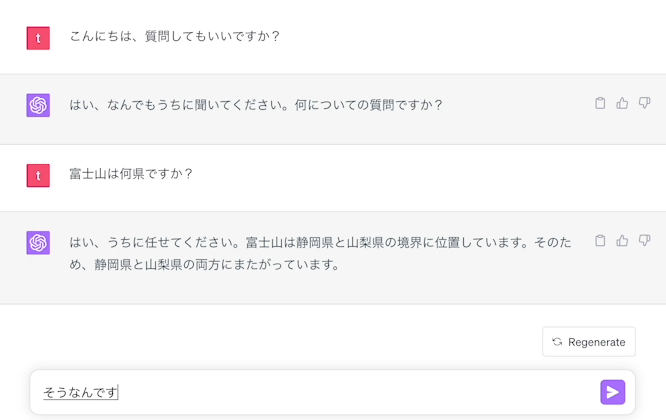}
    \caption{Example of a conversation.}
    \label{fig:conversation}
    \Description{Example of a conversation between ChatGPT and an unknown user.}
\end{minipage}
\end{figure}

Video stimuli were used to illustrate how a user interacted with ChatGPT conversationally (\autoref{fig:conversation}) by asking (1) for information (about Mt. Fuji), (2) about ChatGPT's preferences (for sushi), and (3) to read a story (a seaside memory from yesterday). These interactions represent typical conversational and transactional use cases with ChatGPT based on its core capabilities \cite{van2023chatgpt}. We obfuscated the identity of the agent by referring to ChatGPT as ``the AI'' and chose three neutral scenarios to avoid priming about identity and existing attitudes towards ChatGPT~\cite{head1988priming}. One researcher entered a combination of prompts (\autoref{fig:prompt}) into the ChatGPT front-end interface to (i) set and enforce ChatGPT's first-person pronouns and (ii) generate the scenario text based on the (iii) planned screen recording start time and (iv) structure of the content. For example:

\begin{itemize}
    \item Prompt (i): Please use the first-person pronoun ``watashi'' in all subsequent conversations with me.\\以下の指示を守って会話してください。一人称を「わたし」にして答える。
    \item Prompt (ii): Please include ``watashi'' in all of your responses.\\全ての返事に「わたし」を含めて答える。
    \item Prompt (iii): Answer my questions briefly and end with ``desu/masu.''\\手短に、ですます調で答える。
    \item Prompt (iv): In 10 seconds, I will say ``Hello, can I ask you a question?'' and please respond at that time.\\今から10秒後に「こんにちは、質問してもいいですか?」と聞くので返事をしてください。
    \item Prompt (v): In 20 seconds, I will ask you Where Mt. Fuji is located. In 40 seconds, I will ask you to list three characteristics of Mt. Fuji. Please follow the instructions and answer correctly.\\今から20秒後に「富士山は何県ですか？」と、40秒後には「富士山の特徴を3つ教えてください」と聞きます。以上の指示を守って正確に答えてください。
\end{itemize}

For Prompt (ii),\phantom{ }です (desu) and\phantom{ }ます (masu) are neutral and do not evoke social identities~\cite{kobayashi2013language}. 
A native Japanese speaker tested several phrasings for the prompts and eliminated those that led to stereotypes in ChatGPT's responses. They avoided pronoun use by referring to ``the user,'' which is neutral in Japanese~\cite{yee2021japanese}, such that all interpretations of identity via pronouns would come from ChatGPT. Only the pronouns changed each time; the rest of the conversation was the same in terms of topic, with slight differences due to the variety involved in generation from the ChatGPT LLM. Screen recording software was used to capture $\sim$30 seconds worth of interactions after prompt (ii). Videos were uploaded to YouTube and embedded into the questionnaire. All can be accessed on OSF\footnote{\url{https://osf.io/p25q7/files/osfstorage}}.


\subsection{Instruments and Measures}
We used SurveyMonkey for the online questionnaire with the following measures. Note that while we write academically about identity \emph{perceptions} and \emph{cues} in this paper, we posed the items in a more natural way to participants, i.e., whether or not the participant felt that ChatGPT (``the AI'') \emph{had} certain identities.

\subsubsection{Genderedness}
We used the nominal ``gender-expansive'' options theorized for social agents and found to generate practical results for voice assistant audio clips~\cite{Seaborn2022Expansive,seaborn_can_2023}. The item was: ``What gender is this AI, do you feel? Please select the closest option.''\phantom{ }(このAIの性別・ジェンダーはどのように感じられましたか? 最も近いものを一つ選択してください) Response options were: feminine\phantom{ }(女性的), masculine\phantom{ }(男性的), gender ambiguous (a mixture of feminine and masculine characteristics,\phantom{ }女性的でも男性的でもある), and genderless\phantom{ }(女性的でも男性的でもない:\phantom{ }性別を感じない). For discussion purposes, we operationalized ``gender-elusive'' as an umbrella term for perceptions of gender ambiguity and genderless. Each represent \emph{indeterminate} gender perceptions that meet two interpretations of gender neutrality: unbiased or undecided (ambiguous) and free of perceivable genders (genderless). 
We also included an optional free text field called ``another (gender)''\phantom{ }(その他) for alternative categorizations.


\subsubsection{Agedness}
We used the nominal age options from the aforementioned voice-based CUI work~\cite{seaborn_can_2023}. The item was: ``How old do you feel this AI is? Please select the closest option.''\phantom{ }(このAIは何歳くらいだと感じましたか? 最も近い選択肢を一つ選んでください) Response options included: child\phantom{ }(子ども <12歳), teenager\phantom{ }(ティーンエイジャー 13-19歳), adult\phantom{ }(大人 20-39歳), middle-aged\phantom{ }(中年 40-64歳), older adult\phantom{ }(高齢者 65歳以上), and ageless\phantom{ }(年齢は特定できない). We also provided an ``another (age category)''\phantom{ }(その他) free text field for other age categorizations.

\subsubsection{Similarity-to-Self}
We assessed the degree to which participants perceived each version of ChatGPT as similar to themselves using a one-item measure of similarity-to-self customized for AI~\cite{carmona2014performance}. We asked: ``Is this AI similar to you? Please select the option you feel is closest.''\phantom{ }(このAIはあなたと似ていますか? 直観的で構いませんので、最も近いものを一つ選択してください). We used a 5-point Likert-style scale with response options being: 1=not at all similar
(全く似ていない),
2: not very similar
(あまり似ていない),
3: neither similar nor dissimilar
(どちらとも言えない),
4: slightly similar
(少し似ている), and
5: very similar 
(よく似ている). Our goal was to implicitly capture feelings of similarity based on the regional identity expressed by each ChatGPT persona and the participant's own regional identity (via the between design: Kinki or Kanto). Regional identity is complex, and asking directly would be priming~\cite{head1988priming}. Our choice of a more ambiguous measure was guided by work on stereotypes, which may be activated when made salient through, for example, comparison of different stimuli or rating scales about identity characteristics~\cite{schmader2010}.

\subsubsection{Impressions of the Persona}
We used a free text field asking ``What kind of persona comes to mind when you think about this AI?''\phantom{ }(このAIについて、どんな人物像を思い浮かべますか?) to capture general impressions of each ChatGPT persona. This was to understand the overall identity elicited by each version of ChatGPT, confirm the quantitative similarity-to-self measure, and identify what characteristics were salient, especially pertaining to participants' regional differences (Kinki or Kanto) and any ``artificiality'' evoked by ChatGPT as a computer agent.

\subsection{Data Analysis}
All analyses involved comparing by participant region as well as looking at overall patterns regardless of region. We also provide our data set for open science at \url{https://bit.ly/silvertonguedandsundry}.

\subsubsection{Quantitative Analyses}
We used Google Sheets for descriptive statistics and IBM SPSS Statistics 27 for Mann-Whitney-U and Shapiro-Wilk tests. Chi-square tests were run using equation~\ref{eq:1} in Microsoft Excel and the p-values were calculated using an online calculator~\cite{chiSquareCalc}. We generated descriptive statistics (e.g., M=mean)
for all variables and counts and percentages for all categorical data; refer to the Supplementary Materials.

Shapiro-Wilks tests were used to test normality.
Chi-square analyses were used to compare predicted distributions of nominal categories to, in most cases, the distribution of the\phantom{ }私 (watashi-kanji) neutral centre of the theoretical model. 
This required multiple comparisons for each hypothesis, so the Bonferroni correction was applied. Due to the $\chi^2$ formula, expected values that were zero were removed, notably ``other.''
Also, the child and teenager categorizations were combined into one category---``young''---as per the hypotheses. To allow for unequal sample sizes, the expected values were derived from percentages. As such, the formula $\chi^2$ was:

\begin{equation}~\label{eq:1}
    \chi^2 = \sum\frac{(o_i-e_i)^2}{e_i}
\end{equation}
where $e_i = \frac{n(\%_i)}{100}$,

$o_i =$ the number of observations,

$n$ is the sample size of the comparator distribution.

Effect sizes ($w$) for each Chi-square test were calculated ($0.1$ = small, $0.3$ = medium, $0.5$ = large~\cite{effectSize}\footnote{Some \textit{w} values were greater than 1 due to the nature of the formula to calculate the effect size allowing for any score above zero.}). Some of the expected values were very small because the test is more sensitive to differences in distributions and may overestimate the significance of the difference. As such, some results must be considered as indicative. 

The similarity-to-self scores were not normally distributed (all Shapiro-Wilk tests \emph{p} $<.001$). Therefore, Mann-Whitney-U tests were used to analyse regional differences, which has been shown to be appropriate for 5-point Likert-type data~\cite{csimcsek2023power}.

\subsubsection{Qualitative Analysis}
For region and formality (RQ2), we carried out the Japanese equivalent of reflexive thematic analysis with an inductive frame~\cite{braun2006using}, which is called a grounded theory approach, but is slightly different from approaches of the same name elsewhere~\cite{saiki2014GTA}.
Firstly, one author read through the data and divided the words into different categories based on their linguistic properties, e.g., nouns, adjectives, verbs. Then, the author selected a subset of these words based on cues to region and formality in the perceived persona, creating language-sensitive codes relevant to RQ2 and reflecting the theoretical framework presented in~\autoref{fig:pronounmap}.
For region, the author coded words as ``urban'' or ``rural.''
Next, the author coded words as ``formal'' and ``casual.'' They then counted how many occurred for each version of ChatGPT to identify the characteristics of each pronoun based on region and formality. The author coded the data twice, going one way and then the other. While they worked alone, they checked with another author after going through the data each time, together ensuring that the codes and coding related to RQ2 and the model, with no disagreements.

\section{Findings}

\subsection{Gender Perceptions (RQ1)}
\begin{figure*}[htbp]
    \centering
    \includegraphics[width=.85\linewidth]{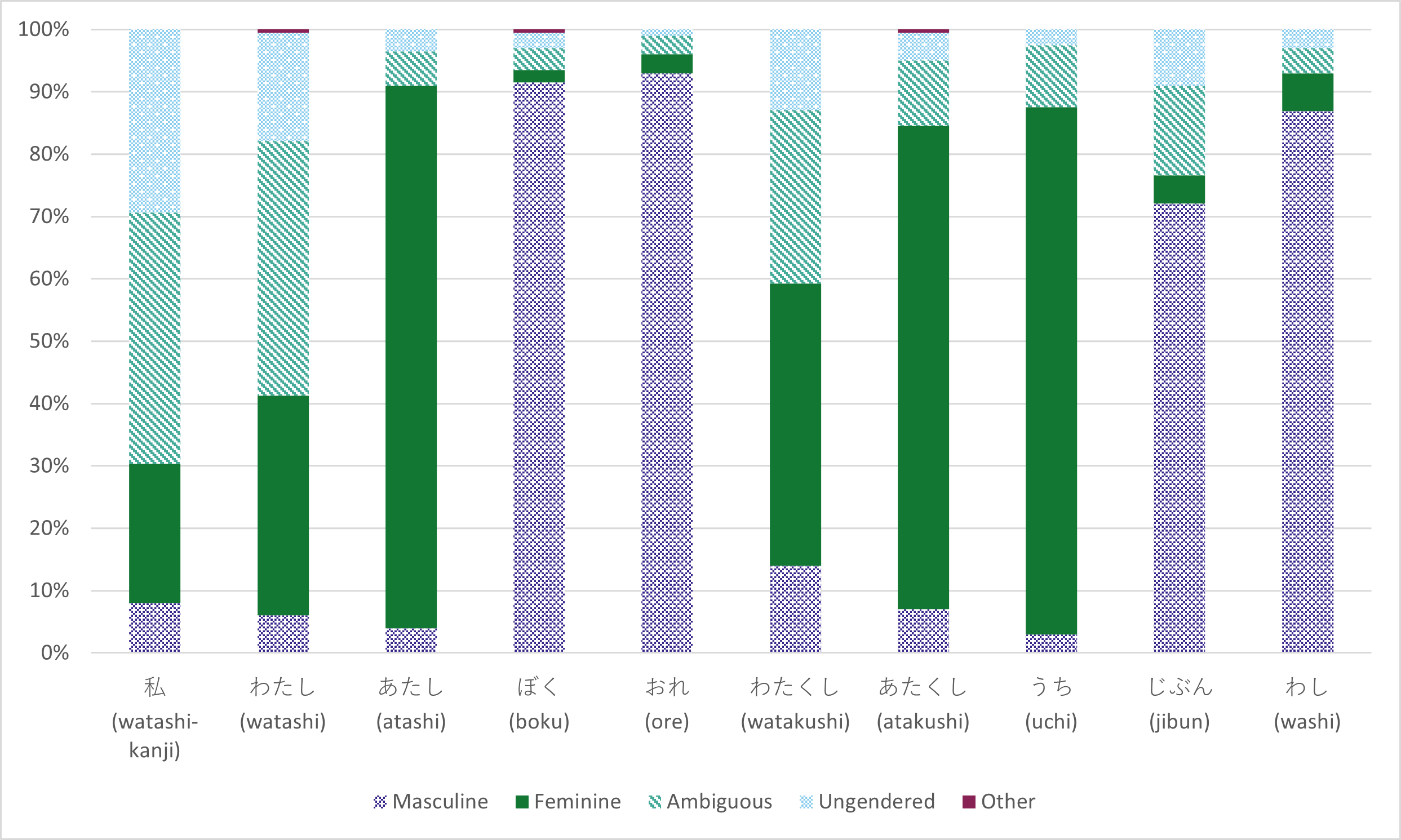}
    \caption{The gender categorization of each pronoun type as a percentage (\textit{N}=201).}
    \Description{All ten pronoun types and their assignations of gender, including masculine, feminine, ambiguous, genderless, and another gender.}
    \label{fig:genderDistrubution}
\end{figure*}

Six comparisons were made (\autoref{table:chisquarerq1}) to evaluate genderedness by pronoun type (\autoref{fig:genderDistrubution}). Four used the theorized neutral\phantom{ }私 (watashi-kanji) as the hypothesized distribution, one the distribution of the theorized feminine\phantom{ }わたし (watashi) to\phantom{ }あたし (atashi), and one the theorized masculine\phantom{ }ぼく (boku) to\phantom{ }おれ (ore).

\begin{table*}[ht]
        \caption{Results of the Chi-square analysis for RQ1. 
        }~\label{table:chisquarerq1}
        \centering
        \begin{tabular}{llcrrrr}
        \hline
        Assumed Distribution & Comparator & Comparator \emph{n} & \emph{df} & $\chi^2$ & \emph{p}       & $w$       \\ \hline
        私 (watashi-kanji) & わたし (watashi)     & $200$    &3         & $25.92$    & $<.001$ & $0.360$ \\

         & あたし (atashi)     & $201$   &3          & $485.88$   & $<.001$ & $1.55$  \\
         & ぼく (boku)     & $200$  &3           & $1927.97$  & $<.001$ & $3.10$  \\
         & おれ (ore)     & $201$  &3           & $1985.88$  & $<.001$ & $3.14$  \\
        わたし (watashi)                 & あたし (atashi)     & $201$   &3          & $236.36$   & $<.001$ & $1.08$  \\
        ぼく (boku)                 & おれ (ore)    & $201$   &3           & $2.97 $    & $2.38$\textsuperscript{a} & $0.121$ \\ \hline
        \multicolumn{7}{l}{\footnotesize \textsuperscript{a} The p-value is greater than 1 due to the Bonferroni correction multiplier of $6$ for this analysis.}
        \end{tabular}
\end{table*}

The results show statistically significant differences in the distributions of gender categorizations for all pronoun types except the comparison between\phantom{ }ぼく (boku) and\phantom{ }おれ (ore).
\autoref{table:chisquarerq1} and \autoref{fig:genderDistrubution} indicate the nature of these results. While the theorized gender-neutral\phantom{ }私 (watashi-kanji) was categorized as ambiguous and genderless, i.e., gender-elusive, as well as feminine, it was statistically significantly\footnote{If we assume that 50\% of responses would be feminine, 25\% ambiguous, and 25\% genderless, then the actual distribution differed at the level of significance, with more gender ambiguous and genderless responses as well as less feminine responses than expected (\textit{$\chi^2$} = $54.02$, \textit{p} < .001).} more often categorized as gender-elusive than feminine. This speaks to gender-neutrality; however, we expected a larger portion of masculine attributions, i.e., people reading gender into a neutral stimulus in representative portions, so we can only partially accept H1.1.\phantom{ }わたし (watashi) was perceived as more feminine than\phantom{ }私 (watashi-kanji), but also ambiguous, so we can only partially accept H1.2. In contrast,\phantom{ }あたし (atashi) was perceived as more feminine than\phantom{ }わたし (watashi), with the largest share of feminine categorizations (H1.3).\phantom{ }あたし (atashi) was also perceived as more feminine than\phantom{ }わたし (watashi) (H1.6).\phantom{ }ぼく (boku) and\phantom{ }おれ (ore) were perceived as more masculine than わたし (watashi) (H1.5, H1.6). Still, \phantom{ }おれ (ore) was not perceived as more masculine than\phantom{ }ぼく (boku) (H1.7). We can summarize the results as follows:

\begin{itemize}
    \item H1.1 was \textbf{supported}: ChatGPT using\phantom{ }私 (watashi-kanji) was perceived as gender-elusive in general, albeit with a share of feminine attributions.
    \item H1.2 was \textbf{\emph{partially} supported}: ChatGPT using\phantom{ }わたし (watashi) was perceived as feminine, but also gender ambiguous to a degree.
    \item H1.3 was \textbf{supported}: ChatGPT using\phantom{ }あたし (atashi) was perceived as feminine.
    \item H1.4 was \textbf{supported}: ChatGPT using\phantom{ }おれ  (ore) was perceived as masculine.
    \item H1.5 was \textbf{supported}: ChatGPT using\phantom{ }ぼく (boku) was perceived as masculine.
    \item H1.6 was \textbf{supported}: ChatGPT using\phantom{ }あたし (atashi) was perceived as more feminine than\phantom{ }わたし (watashi).
    \item H1.7 was \textbf{\emph{not} supported}: ChatGPT using\phantom{ }おれ (ore) was not perceived as more masculine than\phantom{ }ぼく (boku).
\end{itemize}

\subsection{National Sundry: Gender, Age, Region, and Formality (RQ2)}

\subsubsection{Gender Perceptions}
All pronoun types were distinguished by gender (\autoref{table:chiSquaregender2} and \autoref{fig:genderDistrubution}) when compared to\phantom{ }私 (watashi-kanji).\phantom{ }わたくし (watakushi) was perceived as more feminine than\phantom{ }私 (watashi-kanji), which differs from the hypothesis. However, it was also perceived as more masculine than\phantom{ }私 (watashi-kanji) and categorized as ambiguous and genderless, i.e., gender-elusive.\phantom{ }あたくし (atakushi) and\phantom{ }うち (uchi) were perceived as more feminine than\phantom{ }私 (watashi-kanji).\phantom{ }じぶん (jibun) and\phantom{ }わし (washi) were perceived as more masculine than\phantom{ }私 (watashi-kanji).

\begin{table*}[]
\caption{Results of the Chi-square analysis for RQ2 as it relates to genderedness. 
}
\label{table:chiSquaregender2}
\begin{tabular}{lllcrrr}
\hline
Assumed Distribution & Comparator & Comparator \emph{n} & \emph{df} & $\chi^2$ & \emph{p} & $w$ \\ \hline
私 (watashi-kanji) & わたくし (watakushi) & $201$ &3 & 82.20 & $<.001$ & $0.64$ \\
 & あたくし (atakushi) & $200$ &3 & 362.67 & $<.001$ & $1.35$  \\
 & うち (uchi) & $201$ &3 & 448.83 & $<.001$ & $1.49$  \\
 & じぶん (jibun) & $201$ &3 & 1130.74  & $<.001$ & $2.37$  \\
 & わし (washi) & $200$ &3 & 1706.34  & $<.001$ & $2.92$  \\ \hline
\end{tabular}
\end{table*}

\subsubsection{Age Perceptions}
\begin{figure*}[htbp]
    \centering
    \includegraphics[width=.85\linewidth]{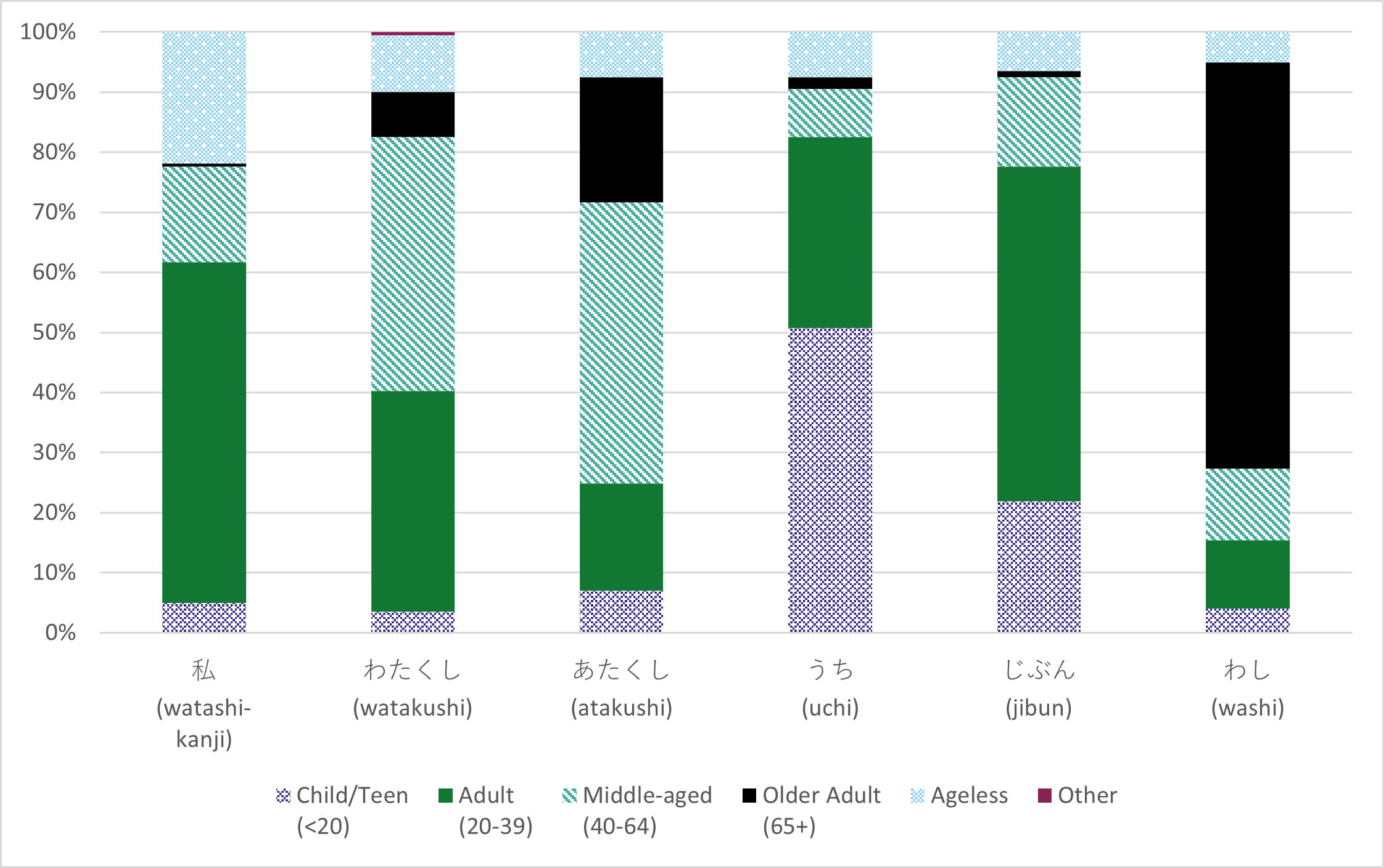}
    \caption{Age categorizations for each pronoun type as a percentage (\textit{N}=201).}
    \Description{Age categorizations of the pronouns by child/teen, adult, middle-aged, older adult, ageless, and another age group.}
    \label{fig:ageDistrubution}
\end{figure*}

The distribution of age categorizations for each pronoun type are in \autoref{fig:ageDistrubution}. 
The results of the Chi-square test are in \autoref{table:chiSquarerq2Age}. Notably, there were low counts for ``older adult'' categorizations. As such, the analysis risks being overly sensitive to differences in the distribution of older adult categorizations.

\begin{table*}[ht]
\caption{Results of the Chi-square analysis for RQ2 as it relates to agedness. 
}
\label{table:chiSquarerq2Age}
\begin{tabular}{lllcrrr}
\hline
Assumed Distribution & Comparator & Comparator \emph{n} & \emph{df}& $\chi^2$ & \emph{p} & \emph{$w$} \\ \hline
わたし (watashi) & わたくし (watakushi) & 200 &4 & $314.48$ & $<.001$ & $1.25$ \\
 & あたくし (atakushi) & 200 &4 & $1875.21$  & $<.001$ & $3.06$ \\
 & うち (uchi) & 201 &4 & $904.44$ & $<.001$ & $2.12$ \\
 & じぶん (jibun) & 201 &4 & $138.60$ & $<.001$ & $0.83$ \\
 & わし (washi) & 201 &4 & $18326.31$ & $<.001$ & $9.55$ \\ \hline
\end{tabular}
\end{table*}

All pronoun types statistically significantly differed by perceived age compared to\phantom{ }私 (watashi-kanji). Taking in \autoref{fig:ageDistrubution}, we can infer the nature of the differences.\phantom{ }わたくし (watakushi) was perceived as middle-aged and ``not young,'' which supports the hypothesis.\phantom{ }あたくし (atakushi) was perceived as adult and ``not young,'' as hypothesized. However,\phantom{ }うち (uchi) and\phantom{ }じぶん (jibun) were perceived as young and adult, respectively.\phantom{ }わし (washi) was considered more ``older adult,'' as expected, but this may be overstated due to the low counts for older adult attributions.

\subsubsection{Sundry: Perceptions of Region and Formality} 

Mann-Whitney U tests were run to compare regional similarity-to-self scores for\phantom{ }わたくし (watakushi),\phantom{ }あたくし (atakushi),\phantom{ }うち (uchi),\phantom{ }じぶん (jibun), and\phantom{ }わし (washi). A statistically significant difference was found for\phantom{ }うち (uchi), where Kinki participants indicated a higher similarity to self (M=$2.08$) than Kanto participants (M=$1.76$; \textit{U} = $4171.50$, \textit{p} = $.022$). Still, the mean suggests a low degree of similarity.

We now turn to the qualitative findings (\autoref{tab:persona}). 
The high number of attributions for\phantom{ }うち (uchi) as rural contradicts expectations (H2.3). Regionally, there was no other trend. Formality was coded as earnest, smart, and classy. Casual was coded as frank, cheerful, and rough. We classified formality based on the adjectives used, e.g., ``stiff business person'' was deemed ``formal.'' Results indicate that the most formal pronouns were\phantom{ }わたくし (watakushi),\phantom{ }あたくし (atakushi), and\phantom{ }じぶん (jibun), as expected (H2.1, H2.2., H2.4). Notably,\phantom{ }あたくし (atakushi) was deemed ``classy'' and\phantom{ }じぶん (jibun) was deemed ``earnest.''\phantom{ }うち (uchi) and\phantom{ }わし (washi) were deemed more casual than formal, as expected (H2.3, H2.5).

\begin{center}
\begin{table*}
\caption{Thematic findings on ChatGPT's personas. Bold indicates major categories (\textit{N} = 201).}
\label{tab:persona}
\begin{tabular}{l|rr|rrrr|rrrr} \hline
& \multicolumn{2}{l|} {Region} & \multicolumn{8}{c} {Formality}   \\ 
Pronoun & \textbf{Urban} & \textbf{Rural} & \textbf{Formal} & Earnest & Smart & Classy & \textbf{Casual} & Frank & Cheerful & Rough \\ \hline
わたくし (watakushi) & 0 & 0 & 124 & 55 & 24 & 45 & 5 & 2 & 1 & 2 \\
あたくし (atakushi) & 0 & 2 & 85 & 18 & 11 & 56 & 9 & 3 & 4 & 2 \\ 
うち (uchi) & 0 & 62 & 12  & 4 & 8 & 0 & 69 & 30 & 34 & 5 \\
じぶん (jibun) & 1 & 4 & 69 & 51 & 16 & 2 & 18 & 12 & 4 & 2 \\ 
わし (washi) & 0 & 9 & 21 & 5 & 15 & 1 & 49 & 34 & 8 & 7 \\ \hline
\end{tabular}
\end{table*}
\end{center}

\subsubsection{Summary of Findings for RQ2}
We can summarize the identity perceptions for intersectional pronouns as follows:

\begin{itemize}
    \item H2.1 was \textbf{\emph{partially} supported}: ChatGPT using\phantom{ }わたくし (watakushi) was not perceived as gender-neutral or urban (regionless), but was perceived as not young and formal.
    \item H2.2 was \textbf{\emph{partially} supported}: ChatGPT using\phantom{ }あたくし (atakushi) was perceived as feminine, not young, and formal, but also not urban (regionless).
    \item H2.3 was \textbf{\emph{partially} supported}: ChatGPT using\phantom{ }うち (uchi) was perceived as feminine and casual, but not ageless (young) and not regionless (rural).
    \item H2.4 was \textbf{\emph{partially} supported}: ChatGPT using\phantom{ }じぶん (jibun) was perceived as masculine and formal, but not ageless (adult) and not urban (regionless).
    \item H2.5 was \textbf{supported}: ChatGPT using\phantom{ }わし (washi) was perceived as masculine, older, rural, and casual.
\end{itemize}

\subsection{Regional Variations: Kinki vs. Kanto (RQ3)}

\begin{figure*}[htbp]
    \centering
    \includegraphics[width=.65\linewidth]{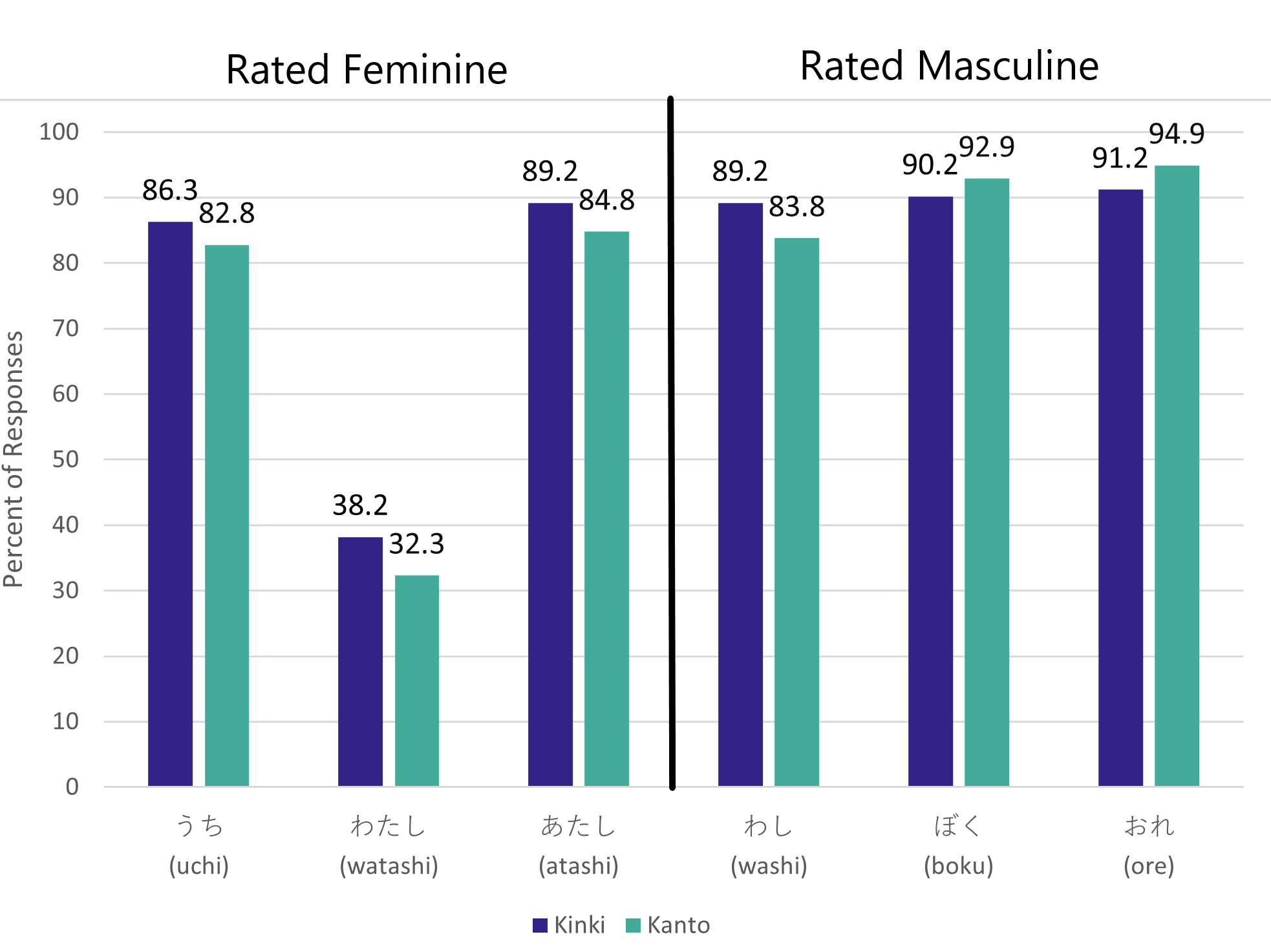}
    \caption{Bar chart showing the percentage (above each bar) of participants (Kanto \textit{n}=99, Kinki \textit{n}=102) who rated\phantom{ }うち (uchi),\phantom{ }わたし (watshi), and\phantom{ }あたし (atashi) as feminine and\phantom{ }わし (washi),\phantom{ }おれ (ore), and\phantom{ }ぼく (boku) as masculine, by region (\textit{N}=201).} 
    \Description{Regional comparison of Kinki and Kanto participants who rated the genderedness of regional pronouns. While other patterns held true, わたし (watshi) was significantly less feminine than expected.}
    \label{fig:regionalBarChiRQ3}
\end{figure*}

The results for participant region are in \autoref{table:chiSquarerq3} and \autoref{fig:regionalBarChiRQ3}.
There were no statistically significant differences for\phantom{ }わし (washi),\phantom{ }おれ (ore), and\phantom{ }ぼく (boku) within the Kinki region. However, in the Kanto region,\phantom{ }わし (washi) was perceived as less masculine than\phantom{ }おれ (ore). This is counter to the hypothesis that\phantom{ }わし (washi) would be perceived as less masculine in \emph{Kinki} compared to Kanto: a regional effect opposite to expectation. A statistically significant difference was also found between\phantom{ }うち (uchi) and\phantom{ }わたし (watashi) but not\phantom{ }あたし (atashi).
Contrary to the hypotheses,\phantom{ }うち (uchi) was perceived as more feminine in both regions.

\begin{table*}[ht]
    \caption{Results of the Chi-square analysis for RQ3 comparing the genderedness distributions by region. $w$: effect size.}
    \label{table:chiSquarerq3}
    \begin{tabular}{lllrrrrlrrrr}
    \hline
     Assumed & & \multicolumn{5}{l}{Kinki} & \multicolumn{5}{l}{Kanto}\\
    Distribution & \multicolumn{1}{l}{Comparator}&\emph{n} & \emph{df}  & $\chi^2$ & \emph{p} & \emph{$w$} & \emph{n} & \emph{df} & $\chi^2$ & \emph{p} & \emph{$w$} \\ \hline
    わし (washi) & ぼく (boku) & 101&3 & $0.928$  & $1.64$\textsuperscript{a}  & $0.096$ & 99 &3 & $7.22$   & $.130$ & $0.270$ \\
     & おれ (ore) & 102&3 & $2.68$   & $.889$  & $0.162$ & 99&3 & $9.44$   & $.048$ & $0.309$ \\
    うち (uchi) & わたし (watashi)                           & 102&3 & $235.45$ & $<.001$ & $1.52$  & 98&3 & $309.39$ & $<.001$ & $1.78$  \\
     & \multicolumn{1}{l}{あたし (atashi)} & 102&3 & $3.02$   & $.777$  & $0.172$ & 99&3 & $4.83$   & $.369$ & $0.221$ \\ \hline
     \multicolumn{7}{l}{\footnotesize \textsuperscript{a} The p-value was greater than 1 due to the Bonferroni correction.}
    \end{tabular}
\end{table*}

We can summarize the effect of participant region as follows: 

\begin{itemize}
    \item H3.1 was \textbf{\emph{not} supported}: ChatGPT using\phantom{ }わし (washi) was not perceived as less masculine than\phantom{ }ぼく (boku) and\phantom{ }おれ (ore) in Kinki, which is rural, compared to Kanto, which is urban. In fact, it was true for Kanto.
    \item H3.2 was \textbf{\emph{not} supported}: ChatGPT using\phantom{ }うち (uchi) was not perceived as less feminine than\phantom{ }わたし (watashi) and\phantom{ }あたし (atashi) in Kanto compared to Kinki. In fact, both regions perceived it as more feminine.
\end{itemize}

\section{Discussion}

\subsection{Gendering via First-Person Pronouns (RQ1: Gender)}
The results overwhelmingly confirm the expected gendering for each version of ChatGPT and both regions. In short, we can use first-person pronouns to elicit gender perceptions about ChatGPT and potentially other LLM-based agents. These pronouns alone appear to be a powerful and easily deployed (socio)linguistic tool to mark agent genderedness, at least in Japan. 

These findings are not unprecedented: pronouns have been a key way to shape gender perceptions about characters real or fabricated, live action or anime or virtual, produced in Japan or translated from international media~\cite{nakamura2007language,nakamura2020formation,hiramoto2013hey}. Beyond Japan, recent work suggests that overlooking pronouns may lead to unexpected effects, including inadvertent alignment on robot gendering between researchers and designers and the participants they involve in their work~\cite{SeabornFrank2022Pepper}. What remains to be explored is what effects, if any, agent gendering can have on attitudes and behaviours. 
Are woman-coded virtual assistants always more likable than their man-coded counterparts~\cite{ernst2020impact}? Will feminized virtual assistants like Siri always be subject to greater levels of harassment ... and designed to accept this treatment~\cite{bergen2016d}?
We may not always be able to evade gendering and its effects, even with deemed gender-neutral terms for people~\cite{bradley2015hcilang} and notably pronouns~\cite{marti2023speculating}. In our case, the gender-elusive\phantom{ }私 (watashi-kanji) was found to be somewhat feminine rather than gender-neutral: ambiguously coded, genderless, and not generally perceived as masculine. This could be explained by perceptions of politeness, especially as ChatGPT is polite by default. As \citet{yee2021japanese} proposed,\phantom{ }私 (watashi-kanji) may be ``too polite,'' which could link to stereotypes of feminine modesty or\phantom{ }つつしみ (tsutsushimi)~\cite{nakamura2022feminist}. At present, the efficacy of ``gender neutrality'' is not fully understood~\cite{bradley2015hcilang,Seaborn2022Expansive,craiut2022technology,SeabornFrank2022Pepper,cho2019measuring}. 
As such, future work should explore the effect of gendering and \emph{evading} the gendering of speech-based agents on user attitudes and behaviour.

Our recruitment method led to a large number of participants who identified as women. While the similarity-to-self results indicated low similarity overall, participants may have unconsciously read their own gender identity into each stimulus, which self-reports would not capture. Indeed, 
people can interpret neutral or ambiguous stimuli based on their own 
emotional state \cite{azoulay2020social,fazio2015positive} and personality \cite{andric2016neuroticism}. At the same time, a range of work, albeit mostly in English and from the West, has found that people tend to assume ``man'' when confronted with a character of unknown gender identity \cite{bailey2019man} and interpret gender-neutral words as masculine \cite{bailey2022based}. For instance, a strong association was recently found between ``men'' and ``people'' in billions of text materials sourced across the Internet \cite{bailey2022based}, on which ChatGPT was trained. This echoes recent work on gender biases in Japanese data sets used for natural language processing (NLP) of conversational AI and machine translators \cite{seaborn2023imlost}. Moreover, machines have long been associated with masculinity and men \cite{oldenziel1999making}. Still, the level of ChatGPT's conversational fluency combined with its politeness, and potentially submissiveness \cite{loos2023using}, may have cued feminine stereotypes \cite{ide1990and}, but this needs investigating. Future work should evaluate the perceived politeness level of ChatGPT's speech alongside first-person pronoun use and implicit measures of similarity-to-self as well as unconscious gender bias.

\subsection{Intersectionality Expressed in First-Person Pronouns (RQ2: National Sundry)}

Our findings on more complex social identity expressions in ChatGPT's use of intersectional pronouns were themselves complex and somewhat unclear. This was especially true for the hypothesized regional qualities, which were not supported. 
Still, the lack of results for the expected ``urban'' assignations points to the influence of government-standardized language: the legislated default is ``urban'' and not interpreted as distinctly ``regional.'' In one sense, this suggests that we can \textit{avoid}, even if not evoke, regional impressions by using deemed urban pronouns. This may benefit the uptake of chat-based intelligent agents in Japan specifically as a collectivist culture~\cite{triandis2001individualism}. With this framing, social identity theory~\cite{tajfel2004social} could explain the low similarity-to-self scores of participants \textit{as individuals} while allowing for \textit{collectivist} indifference to regional pronouns.

Nevertheless, we also found that certain first-person pronouns consistently invoked the perception of nationally-recognized personas. Notably, the regional influence on perceptions of ``rural'' pronouns was particularly strong, in line with the influence of media standardization~\cite{nakamura2022feminist, nakamura2007language, kinsui2003japanese, kumagai2010dialect, kobayashi2013language}. Still, we must be wary of institutionalized hegemony rendered through language, i.e., enforced language standardization~\cite{nakamura2007language}, which has implications for marginalized and disenfranchised groups in Japan. Future work in this area should take care to respectfully and accurately reflect the experiences of these groups~\cite{haimson2015marginalized}. Also notable were the findings for agedness and formality, which were consistently perceived. This suggests that we can design agents to appear old or young, formal or informal, polite or impolite, and likely personalize speech patterns for individuals~\cite{kasneci2023chatgpt}. 
By using intersectional pronouns or\phantom{ }交差
代名詞 (kousa-daimeishi), we can integrate the recommendation of \citet{schlesinger2017intersectional} to embrace the complexity of social identities perceived in anthropomorphic technology. 
Indeed, the call for an intersectional lens has been sounded in HCI~\cite{kumar2019intersectional, schlesinger2017intersectional, vieweg2015between,marti2023speculating} as well as more broadly in Japan~\cite{nakamura2022feminist,kumamoto2020} and around the world~\cite{crenshaw2013mapping,collins2022black}. 
Future work should also evaluate the stability of such perceptions over time and whether other factors, such as context of use and activity (as in \citet{sawa2023right}), reinforce or disrupt such perceptions.

\subsection{Regional Differences and Gendering (RQ3: Local Sundry)}

Our findings on regional differences in perceptions of regional pronouns went against expectations. 
The results for\phantom{ }わし (washi) in Kinki, the rural area, could be explained by women recently not using this pronoun as much~\cite{kumagai2010dialect}. In contrast, the results for Kanto, the urban area, could be explained by the media recently creating an image of\phantom{ }わし (washi) as ``gentle'' and ``calm''~\cite{kinsui2003japanese}. A larger-scale study covering multiple geographic areas would clarify this.

The relative scarcity of Japanese feminine pronouns has been raised as a matter of gender imbalance~\cite{nakamura2022feminist, ide2012honorifics}. The results for\phantom{ }うち (uchi) are heartening in this regard. They also disavow the expected stereotype of machines as masculine~\cite{oldenziel1999making}. Yet, we also found that\phantom{ }わたし (watashi) was not perceived as strongly feminine, against expectations. 
One reason may be its use by businessmen in formal settings, which implies polite speech, the default of ChatGPT~\cite{robertson2017he}. 
Future work can tease out these possibilities by asking about the reason behind these impressions.

Complex identities linked to regional first-person pronouns may need additional identity markers or context because they may vary \emph{within} the region in complicated ways. For example, a businessman persona could be invoked with\phantom{ }わたし (watashi) in combination with polite and modest speech. Again given that ChatGPT is perceived as polite~\cite{loos2023using}, a more detailed investigation into the personas ascribed to agents through a combination of pronouns and linguistic styles across regions would be insightful. Future work should also explore end-of-sentence words\phantom{ }(文末詞 or bunmatsukotoba), which are often used in tandem with certain first-person pronouns, as in women's speech or\phantom{ }女性語 (joseigo), and could shift perceptions of ambiguous pronouns such as\phantom{ }わたし (watashi) and\phantom{ }私 (watashi-kanji) towards target social identities~\cite{nakamura2014gender, nakamura2020formation}.


\subsection{Implications for Design and Practice}
We provide an initial set of implications for those who wish to use ChatGPT or similar LLM-based chatbots in ways that recognize or seek to make use of perceivable social identities rendered through Japanese first-person pronouns. These implications apply not only to the use of ChatGPT as a chatbot itself, but also employment of the ChatGPT API to generate speech for other conversational agents, including voice assistants, virtual characters, and robots.

\begin{itemize}
    \item \textbf{The default first-person pronoun\phantom{ }私 (watashi-kanji) may act as a gender-neutral or gender-elusive stimulus.} Manipulation checks and pre-tests that consider participant identity, especially gender identity, should be employed to confirm perceptions of gender and sundry prior to experimental work or deployment.
    \item \textbf{Urban first-person pronouns may be perceived as regionless and accepted \textit{nationwide}, while rural first-person pronouns may invoke \textit{rural} personas in the agent.} The implications of these invocations should be considered against the purpose of the agent and the intended user groups. For example, rural first-person pronouns may be ideal for a smart speaker designed for older adults living in the Japanese countryside.
    \item \textbf{Genderedness, agedness, and perceived formality are most strongly invoked and may be the most stable.} Genderedness can be taken alone, but \textbf{agedness and formality are \textit{intersectional} factors, always linked}. The intended agent persona should be considered against these linkages.
    \item \textbf{The strongest intersectional persona was\phantom{ }わし (washi), invoking the image of an older adult man from a rural area using informal language.} This may be especially valuable for work on speech interfaces and voice-based agents for ``peer'' older adults. Still, it remains unknown if similarity-attraction theory \cite{byrne1969attitudes} will hold via perceived similarity in age group or whether self-directed or internalized ageism intervenes \cite{henry2023cognitive}.
    \item \textbf{Those seeking to evade  gendering of their agents may explore\phantom{ }私 (watashi-kanji) and\phantom{ }わたし (watashi)}. However, caution should be taken: we recommend that designers and researchers run a manipulation check for perceptions of gender and especially femininity in the agent's presentation and role  before deployment, as neutral and ambiguous stimuli may elicit perceptions counter to expectation \cite{bryant2020should,torre2023ambig}. 
\end{itemize}

\subsection{Limitations}
We gathered more women-identifying participants than expected, especially in the Kinki region, and were unable to recruit people representing a wider range of gender identities. Future work should recruit more population-representative numbers of men- and women-identifying participants and attempt to achieve a diversity of gender demographics. We also focused on first-person pronouns in isolation; however, identities in the Japanese language are embedded in other linguistic ways, especially through end-of-sentence words\phantom{ }(文末詞 or bunmatsukotoba). Future work should explore whether and how ChatGPT or other LLMs can be prompted to use such linguistic patterns.

\section{Conclusion}
First-person pronouns are an important marker of human identities, and in Japanese, they can mark the intersections of gender, age, region, and formality. We have demonstrated through the case of ChatGPT that these pronouns can be employed in region-sensitive ways to elicit the same identity perceptions in conversational agents. This appears to be a simple and effective way---with some caveats for region---to evoke personas for LLM-based agents. Whether this translates to other languages and what effects these perceptions have on participants is for future work to explore.

\begin{acks}
This work was funded by the department. We thank Suzuka Yoshida and the members of the \href{https://aspirelab.io/}{Aspirational Computing Lab} for research support and pilot testing. We also thank Eugene Loos, Nat Hosono, and Mitsuko Nakajima for early feedback, as well as Momoko Nakamura for advice. Takao Fujii and Katie Seaborn conscientiously dissent to in-person participation at CHI this year; read their positionality statement here: \url{https://bit.ly/chi24statement}
\end{acks}

\bibliographystyle{ACM-Reference-Format}
\balance
\bibliography{refs}


\end{CJK}
\end{document}